\documentclass{elsart}
\journal{Nuclear Physics A}

\usepackage{amsmath} 

\usepackage{graphicx}
\usepackage{epsfig}

\usepackage{amssymb}

\usepackage{lineno}

\begin{document}

\begin{frontmatter}



\title{Production of light flavour hadrons at intermediate and high $p_{\rm{T}}$ in pp, p--Pb and Pb--Pb collisions measured with ALICE}

\author{Michael Linus Knichel (for the ALICE Collaboration)}
\address{Physikalisches Institut, Ruprecht-Karls-Universit{\"a}t Heidelberg, Im Neuenheimer Feld 226, 69120 Heidelberg, Germany}
\ead{knichel@physi.uni-heidelberg.de}

\begin{abstract}
Light flavour transverse momentum spectra at intermediate and high $p_{\rm T}$  provide an important baseline for the measurement of perturbative QCD processes in pp, for the evaluation of 
initial state effects in p--Pb, and for investigating the suppression from parton energy loss in Pb--Pb collisions.
The measurement  of the nuclear modification factor $R_{\rm{pPb}}$  for inclusive charged particles is extended up to 50 GeV/$c$ in  $p_{\rm T}$  compared to our previous measurement  and remains consistent  with unity up the largest momenta.
Results on $R_{\rm{pPb}}$ of charged pions, kaons and protons that cover up to 14 GeV/$c$ in  $p_{\rm T}$  are presented and compared to $R_{\rm{pPb}}$ of inclusive charged particles.
On the production of charged pions, kaons, and protons up to $p_{\rm T} = 20$ GeV/$c$ in Pb--Pb collisions final results for $R_{\rm{AA}}$ are presented.
 
\end{abstract}


\end{frontmatter}


\section{Introduction}
In proton-proton collisions particles at intermediate and high $p_{\rm T}$ are produced
 in hard scattering processes with large virtuality. The cross section for these processes can be calculated with perturbative QCD (pQCD) approaches.
Measurements of transverse momentum spectra provide a test of pQCD calculations which include non-perturbative effects in the parton distribution functions and fragmentation functions. 
 Phenomenological approaches used in Monte Carlo event generators need to be tuned bases on measured data.
Moreover, pp collisions provide a reference (QCD vacuum) to study the effects induced by the QCD medium formed in nucleus-nucleus collisions and by the nuclear initial-state in proton-nucleus collisions.

Initial and final state nuclear effects in p--Pb and Pb--Pb collisions result in particle production that differs from an incoherent superposition of nucleon-nucleon collisions. The difference is quantified by the nuclear modification factor
 \begin{equation}
R_{\rm{AA,pPb}} (p_{\rm{T}}) = \frac{\rm{d}^2 \it{N}^{\rm{AA,pPb}}/\rm{d} \it{y} \rm{d} \it{p}_{\rm{T}}}{\left< \it{N}_{\rm{coll}} \right> \ \rm{d}^2 \it{N}^{\rm{pp}} /\rm{d} \it{y} \rm{d} \it{p}_{\rm{T}}}  = \frac{\rm{d}^2 \it{N}^{\rm{AA,pPb}}/\rm{d} \it{y} \rm{d} \it{p}_{\rm{T}}}{\left< \it{T}_{\rm{AA,pPb}} \right> \ \rm{d}^2 \sigma^{\rm{pp}} /\rm{d} \it{y} \rm{d} \it{p}_{\rm{T}}},
\end{equation} 
where $\left< N_{\rm{coll}} \right>$ is the average number of binary nucleon-nucleon collisions in one Pb--Pb or p--Pb collision and $\left< T \right>$ is the  average   nuclear overlap function. For unidentified charged particles the rapidity $y$ is approximated by the pseudorapidity $\eta$.

Proton-lead collisions allow  the study of  so-called cold nuclear matter (CNM) effects, for instance 
nuclear shadowing (nuclear modification of the parton distribution functions) and $k_{\rm T}$  broadening (due to multiple scattering of the partons prior to the hard scattering).
Collective effects might also be present in p--Pb collisions and are a possible explanation  for  the double-ridge recently observed in di-hadron correlations~\cite{cms:ridge,alice:ridge,atlas:ridge}.
Preliminary results from CMS~\cite{cms:rpaprelim}  and ATLAS~\cite{atlas:rpaprelim}  show an enhancement of particle production above 30 GeV/$c$ in $p_{\rm T}$ compared to expectations from binary collision scaling.
Since CNM effects are present also in nucleus-nucleus collisions they have to be taken into account in the interpretation of Pb--Pb results. 

In collisions of heavy nuclei a hot QCD medium,  the Quark-Gluon Plasma (QGP), is created leading to strong collective effects at low and intermediate $p_{\rm T}$ (most prominently radial flow). 
In addition, parton energy loss in the QGP induces a suppression of the high $p_{\rm T}$ particle yields.

For charged $\pi$, K, p we present preliminary results on $R_{\rm{pPb}}$ ($p_{\rm T} < 14$~GeV/$c$) and final results on $R_{\rm{AA}}$ ($p_{\rm T} < 20$~GeV/$c$). The analysis of charged particles in p--Pb is extended up to $p_{\rm T} = 50$~GeV/$c$ (the first analysis~\cite{alice:rpa} was statistically limited to $p_{\rm T} < 20$~GeV/$c$).

\begin{figure}
\centering
\begin{minipage}[t]{.45\textwidth}
\begin{center}
\includegraphics*[width=\textwidth]{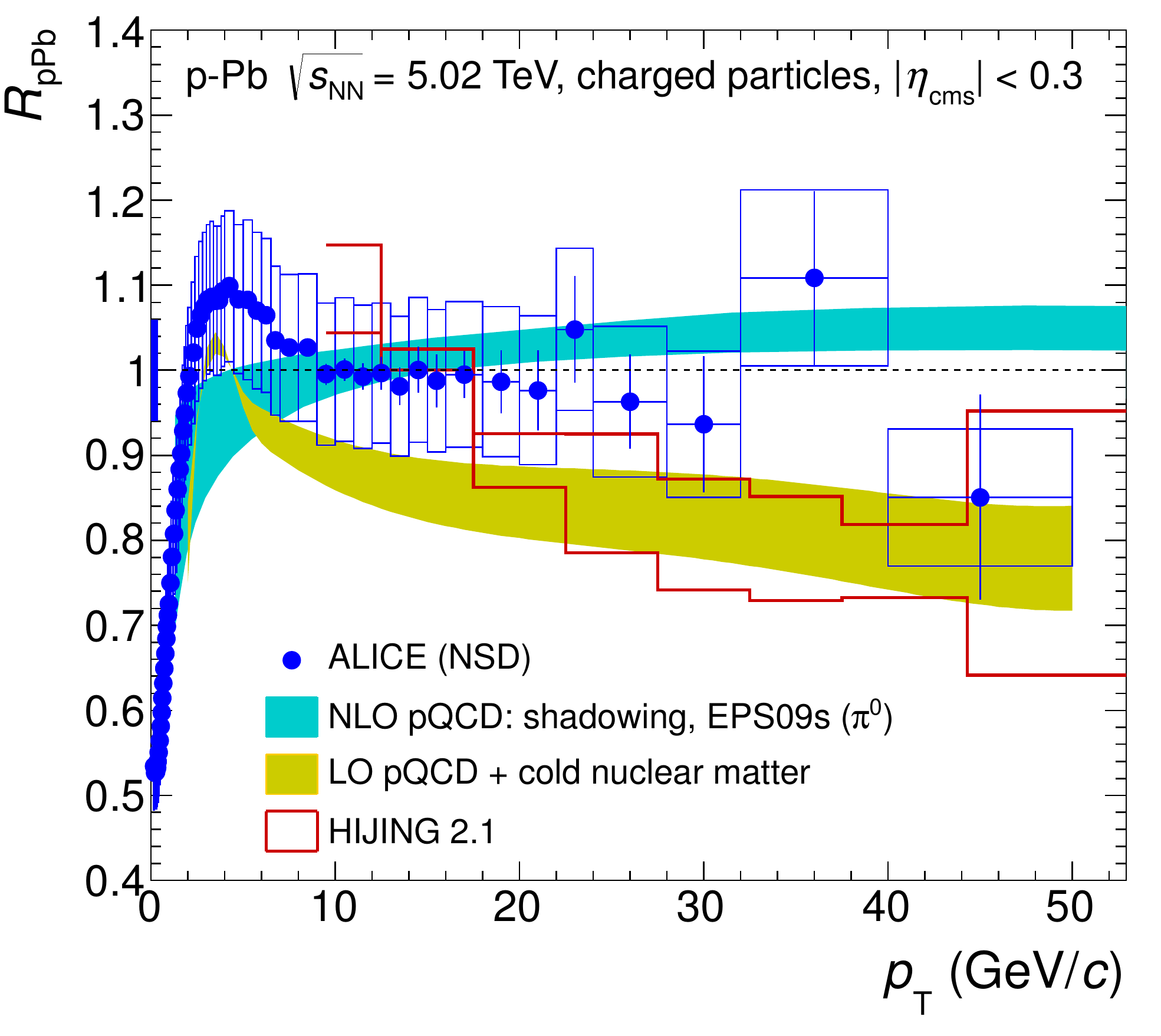}
\caption{
Nuclear modification factor $R_{\rm{pPb}}$ 
of primary charged particles at mid-rapidity $| \eta_{\rm{cms}} | < 0.3$ 
measured in non-single-diffractive (NSD) p--Pb collisions at $\sqrt{s_{\rm{NN}}} = 5.02$~TeV 
\cite{alice:rppb2013} compared to model calculations using NLO pQCD with EPS09s nuclear PDFs~\cite{Helenius:2012wd}, LO pQCD with additional implementation of cold nuclear matter effects~\cite{Kang:2012kc} and HIJING 2.1~\cite{Xu:2012au}.
}
\label{fig:rppb2013models}
\end{center}

\end{minipage}%
\hfill
\begin{minipage}[t]{.45\textwidth}

\begin{center}
\includegraphics*[width=\textwidth]{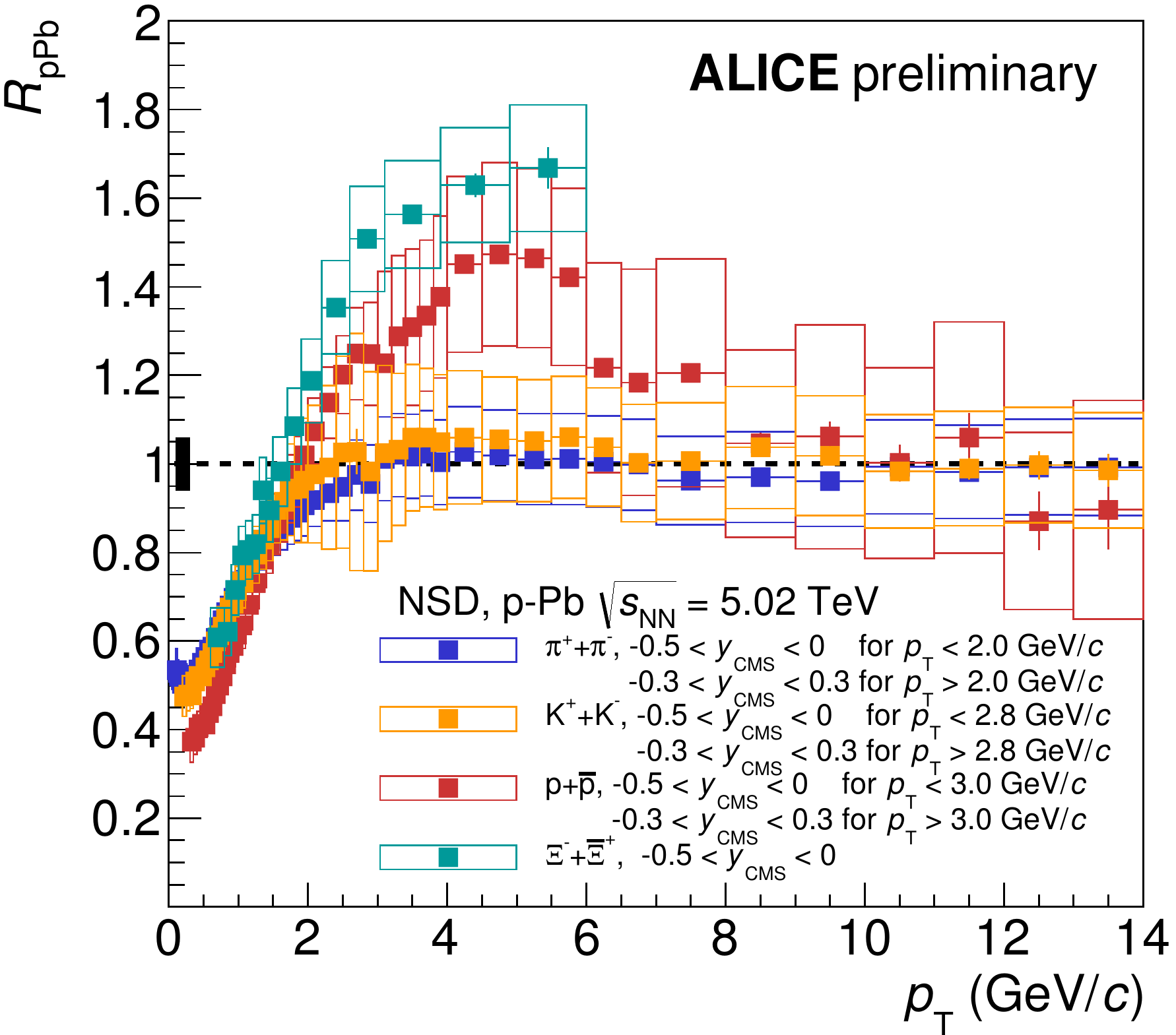}
\caption{
Nuclear modification factor $R_{\rm{pPb}}$ of primary charged $\pi$, K, p and $\Xi$ at mid-rapidity
measured in NSD p--Pb collisions at $\sqrt{s_{\rm{NN}}} = 5.02$~TeV.
}
\label{fig:rppb_all}
\end{center}

\end{minipage}
\end{figure}

\section{Analysis}
The ALICE  experiment~\cite{alice:jinst} is focused on the study of the  hot and dense  QCD medium created in heavy-ion collisions. Among the LHC experiments ALICE has unique particle identification (PID) capabilities at mid-rapidity ($| \eta | < 0.9$).
 The data  presented here were collected with minimum bias triggers from the Silicon Pixel Detector (SPD) and the VZERO scintillators.
Charged tracks and the interaction vertex are reconstructed using the Inner Tracking System (ITS), a six-layer silicon detector, and the Time Projection Chamber (TPC).
Pions with transverse momenta up to 2~GeV/$c$ (kaons up to 2.8~GeV/$c$ and protons up to 3~GeV/c) are identified on a track-by-track  basis  with different detectors providing PID over different momentum ranges (see e.g.~\cite{alice:identifiedraa} for details).
The  differential  energy loss d$E/$d$x$ is measured in the ITS and the TPC and combined with time-of-flight measurements in the TOF detector and the Cherenkov angle from the High Momentum Particle Identification Detector (HMPID).
For transverse momenta above  2-3~GeV/$c$  (currently up to $p_{\rm T} = 20$~GeV/$c$) identified particle spectra are obtained from a statistical analysis  of the energy loss in the TPC in the relativistic rise region.  The inclusive d$E/$d$x$ distribution is fitted -- in intervals of $p_{\rm T}$ and $\eta$ -- by the sum of  four  Gaussians ($\pi$, K, p, e) with their mean and width fixed from  clean samples of identified particles. 
 The  $p_{\rm T}$ distributions are finally obtained as the product of the charged particle spectra and the fractions of $\pi$, K, p from the $\rm d E/\rm d x$ analysis and corrected for the  variations  in acceptance and efficiency.

For the 2013 p--Pb data, with recent improvements in the reconstruction procedure, the transverse momentum resolution achieved by combined ITS-TPC tracking ranges from  $\sigma (p_{\rm T}) / p_{\rm T} \approx 1$\% at $p_{\rm T} = 1$~GeV/$c$ to $\sigma (p_{\rm T}) / p_{\rm T} \approx 3$\% at $p_{\rm T} = 50$~GeV/$c$. 
In the previously reconstructed pp and Pb--Pb data the resolution is $\sigma (p_{\rm T}) / p_{\rm T} \approx   10$\% at $p_{\rm T} = 50$~GeV/$c$.
Tracking and PID performance of the ALICE detector are described in detail in~\cite{alice:performance}.

\section{Results}

Transverse momentum spectra of primary charged particles have been measured in pp collisions at $\sqrt{s} = 0.9$, 2.76 and 7~TeV~\cite{alice:ppreference}. Next-to-leading order (NLO) pQCD calculations~\cite{nlopp} over-predict the measured cross-sections by about a factor 2. A similar discrepancy has been observed  for neutral pions~\cite{alice:pi0pp} and by CMS~\cite{cms:ppreference,cms:raa}. 
Measured jet spectra~\cite{alice:jetspp} are in much better agreement with NLO calculations pointing to the fragmentation functions as main source of the discrepancy.
However, NLO calculations give a reasonable description of the $\sqrt{s}$ dependence of $p_{\rm T}$ distributions and are therefore used to scale the measured charged particle $p_{\rm T}$ spectrum from $\sqrt{s} = 7$~TeV to $\sqrt{s} = 5.02$~TeV, the centre-of-mass energy of p--Pb collisions.
At low transverse momenta ($p_{\rm T} < 5$~GeV/$c$) the reference spectrum is constructed as a bin-by-bin interpolation between the measurements at $\sqrt{s} = 2.76$~TeV and $\sqrt{s} = 7$~TeV, assuming a power law behavior of the cross section with $\sqrt{s}$ for fixed $p_{\rm T}$. For identified hadrons ($\pi$, K, p) the power law interpolation method is used for the full $p_{\rm T}$ range (up to 14~GeV/$c$).

\begin{figure}
\begin{center}
\includegraphics*[width=0.8\textwidth]{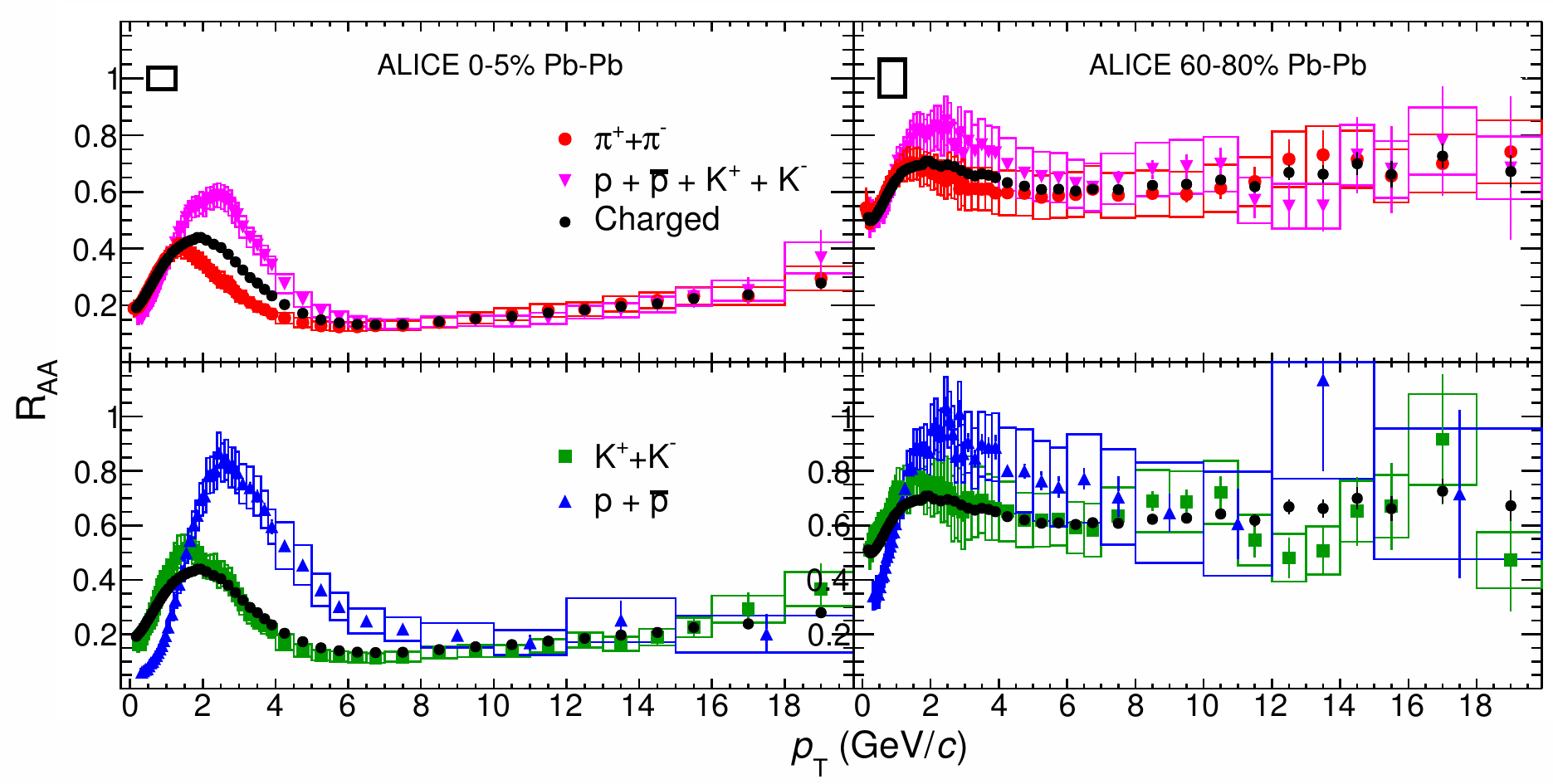}
\caption{
The nuclear modification factor $R_{\rm{AA}}$ of $\pi$, K, p in central (0-5\%) and peripheral (60-80\%) Pb--Pb collisions~\cite{alice:identifiedraa} compared to the $R_{\rm{AA}}$ of inclusive charged particles~\cite{alice:raa2}.}
\label{fig:raa}
\end{center}
\end{figure}

In p--Pb collisions at $\sqrt{s_{\rm{NN}}} = 5.02$~TeV the nucleon-nucleon (NN) centre-of-mass (CM)  system  is moving with $\Delta y_{\rm{cm}} = 0.465$ in the direction of the proton; $y$ and $\eta_{\rm{cms}}$ are in the CM frame.
 As a sign convention we take $y$ and $\eta_{\rm{cms}}$ to be positive along the direction of the proton\footnote{This is opposite to the convention used in~\cite{alice:rpa}.}.
The average nuclear overlap of $\left< T_{\rm{pPb}} \right> = \left< N_{\rm{coll}} \right>/\sigma_{\rm{pp}}^{\rm{INEL}} = 0.0983 \pm 0.0035 \, \rm{mb}^{-1}$ is obtained from a Monte Carlo Glauber simulation with $\sigma_{\rm{pp}}^{\rm{INEL}} = (70 \pm 5)$~mb. 

Figure \ref{fig:rppb2013models}  shows $R_{\rm{pPb}}$  for charged particles at mid-rapidity  $| \eta_{\rm{cms}} | < 0.3$  in p--Pb   collisions~\cite{alice:rppb2013},
compared to three model calculations.  
The next-to-leading order (NLO) pQCD calculation with EPS09s nuclear PDFs~\cite{Helenius:2012wd} describes the charged particle data    for $p_{\rm{T}} > 6$~GeV/$c$. The calculation is for $\pi^0$ and describes well the $R_{\rm{pPb}}$ of charged $\pi$ over the full range ($p_{\rm{T}} < 1.3$~GeV/$c$).
The leading order (LO) pQCD calculation with isospin effect, Cronin effect, cold nuclear matter energy loss and dynamical shadowing separately implemented~\cite{Kang:2012kc} shows a decrease of $R_{\rm{pPb}}$ with $p_{\rm{T}}$ not seen in the data.
In HIJING 2.1~\cite{Xu:2012au}  (shown with different factorization schemes) this decreasing trend is even more pronounced. 

At low $p_{\rm{T}}$ an approximate scaling with the number of participants is observed, corresponding to $R_{\rm{pPb}} \approx 0.57$.
A hint of an enhancement in the Cronin region around $p_{\rm{T}} =$ 2-4~GeV/$c$ is visible. Note that systematic uncertainties are correlated between neighbouring $p_{\rm{T}}$ bins.
Up to the largest $p_{\rm{T}}$ of 50 GeV/$c$ no deviation from binary collision scaling is observed for charged particles,    with an average $\left< R_{\rm{pPb}} \right> = 0.969 \pm 0.056 \rm{(stat.)} \pm 0.090 \rm{(syst.)} \pm 0.06 {(norm.)}$ for $28 < p_{\rm{T}} < 50$~GeV/$c$.
For charged jets $R_{\rm{pPb}}$ is consistent with unity up to 90 GeV/$c$ in $p_{\rm{T}}$~\cite{haake}.
Data presented by CMS~\cite{cms:rpaprelim} and ATLAS~\cite{atlas:rpaprelim} show a rising trend in $R_{\rm{pPb}}$ at large $p_{\rm{T}} > 30$~GeV/$c$ but given the current systematic uncertainties the difference is barely significant.
As part of the difference comes from the reference spectrum, pp data $\sqrt{s} = 5$~TeV would be very useful.

Figure \ref{fig:rppb_all} shows the $R_{\rm{pPb}}$ for charged $\pi$, K, p up to $p_{\rm{T}} = 14$~GeV/$c$ and for $\Xi$ up to 6 GeV/$c$. $\Xi$ baryons are measured in their decay $\Xi^- \rightarrow \Lambda \pi^-$, with the subsequent decay $\Lambda \rightarrow p \pi^-$.
The nuclear modification factor of $\pi$ and K does not differ from the charged particle result within the systematic uncertainties.
At intermediate momenta of $2 < p_{\rm{T}} < 6$~GeV/$c$ a mass ordering with  $R_{\rm{pPb}}^\pi \approx R_{\rm{pPb}}^{\rm{K}} < R_{\rm{pPb}}^{\rm{p}} < R_{\rm{pPb}}^\Xi$  is observed. In Pb--Pb collisions a similar  mass ordering  is observed and attributed to radial flow.
 A notable  Cronin peak is only observed for p and $\Xi$ implying that the small enhancement of $R_{\rm{pPb}}$ for charged particles is driven by the protons.
At larger transverse momenta of $p_{\rm{T}} > 8$~GeV/$c$ the nuclear modification factors of $\pi$, K, p are all consistent with unity.

The nuclear modification factor $R_{\rm{AA}}$ for $\pi$, K, p up to 20 GeV/c~\cite{alice:identifiedraa} is shown in Figure \ref{fig:raa} for central (0-5\%) and peripheral (60-80\%) Pb--Pb collisions at $\sqrt{s_{\rm{NN}}} = 2.76$~TeV with comparison to the charged particle $R_{\rm{AA}}$~\cite{alice:raa2}.
For large transverse momenta ($p_{\rm{T}} > 10$~GeV/$c$) the strong suppression observed  for inclusive charged particles  does not depend on the particle type.

\section{Conclusions}
In p-Pb collisions soft particle production at low $p_{\rm{T}}$ scales approximately with the number of participating nucleons and hard particle production at large $p_{\rm{T}}$ scales with the number of binary nucleon-nucleon collisions, with a smooth transition in between.
This indicates that there are no strong initial or final state nuclear effects at large  $p_{\rm{T}}$. 
At intermediate $p_{\rm{T}}$ the small enhancement (Cronin peak) is  caused by protons.
  Identified light hadrons exhibit a mass ordering of  $R_{\rm{pPb}}$ at intermediate $p_{\rm{T}}$   which is qualitatively similar to that in Pb--Pb collisions.
At $p_{\rm{T}} > 8$~GeV/$c$ $\pi$, K and p $R_{\rm{pPb}}$  is consistent with unity.
In Pb--Pb collisions large suppression is observed for all particle species, with differences from radial flow visible in the $R_{\rm{AA}}$ at low  and  intermediate $p_{\rm{T}}$, while above   $10$~GeV/$c$ the suppression  is common for all species with the p/$\pi$ and K/$\pi$ ratios not  differing  from those in pp collisions.












\end{document}